\documentclass[preprint]{ptephy_v1}

\preprintnumber{XXXX-XXXX} 
\usepackage{hyperref}

\newcommand{\lambdabar}{{\mkern0.75mu\mathchar '26\mkern -9.75mu\lambda}}

\begin{document}
\title{Photon emission by an electron in a constant background field modeling a Lorentz-noninvariant vacuum}


\author{Anatoly V. Borisov}
\affil{Faculty of Physics, M. V. Lomonosov Moscow State University, Moscow 119991, Russia \email{borisov@phys.msu.ru}}





\begin{abstract}
The power and the probability of electromagnetic radiation from an electron in a constant background tensor field violating Lorentz invariance are calculated. 
The case of a background field of the quasielectric type is considered. The angular distribution and the polarization of the radiation are studied. 
Using present experimental constraints on the background field strength, it is shown that the radiation effect can manifest itself
 under astrophysical conditions at ultrahigh electron energy.
\end{abstract}

\subjectindex{B3, B4, B7}

\maketitle

\section{Introduction}
The discovery of the Higgs boson has completed the experimental verification of the Standard Model (SM) of fundamental interactions of elementary particles\cite{pdg}. However, the SM is unable to solve a number of fundamental problems (see, e. g., \cite{lan}): the hierarchy of particle masses and energy scales of strong, electroweak, and gravitational (not described by the SM) interactions, the existence of dark matter and dark energy in the Universe, and others.  In order to solve the above problems, theories generalizing the SM  are developed (see a detailed discussion of a number of them in \cite{nag}). Some of them assume the Lorentz invariance violation (LIV), which can be generated by the effects of quantum gravity, especially significant at ultrahigh energies of the order of the Planck energy ($\sim 10^{19}~\mbox{GeV}$). In the region of relatively low energies, however, the LIV is described by a theory currently under active development called the Standard Model Extension ((SME) of the \cite{sme1,sme2,sme3}, which is a variant of the  effective field theory \cite{wei}.  The SME Lagrangian is the sum of the SM Lagrangian and additional summands --- various combinations of the SM fields with free tensor indices (they ensure the LIV)  convolved with constant coefficients of the corresponding tensor dimension. These coefficients, considered as constant background fields, simulate the complex structure of vacuum due to the new physics not described by the SM.

The effects due to the violation of Lorentz invariance modeled by the interaction with the background axial-vector field (BAVF) have been studied in a number of papers: birefringence of light in vacuum \cite{cfj,sme2}; production of an electron-positron pair by a photon and emission of a photon by an electron and a positron \cite{zhlm1,zhlm2}; synchrotron radiation of an electron  taking into account its anomalous magnetic moment (AMM) and interaction with the BAVF \cite{fzh}; the influence of this interaction on the radiation of the hydrogen-like atom \cite{khzh}; 
the vacuum current generation due to combined interaction of an electron with the BAVF and the electron AMM with a constant magnetic field \cite{bgzh}.

In our paper \cite{bor1}, we calculated the one-loop mass and vertex (at zero momentum transfer) operators of the electron
in a tensor background field of the quasimagnetic type violating Lorentz invariance, and, using the experimental values of the electron mass and charge,
we obtained constraints on the background field strength were obtained. In the papers \cite{bor2,bor3,bor4}, we considered the emission of a photon by an electron moving in the specified background field. The power and probability of the radiation, its angular distribution and polarization were investigated. With use of current experimental constraints on the background field strength, it is shown that the radiation effect can become significant under astrophysical conditions at ultrahigh electron energies.

In the present paper, we consider the electromagnetic radiation of an electron moving in a tensor background field of the quasielectric type, using, as in \cite{bor2,bor3,bor4},
the Lanrangian of the form \footnote{We use a system of units in which $\hbar = c =1$, $\alpha=e^2/4\pi\simeq 1/137$, and a pseudo-Euclidean metric with the signature $(+\,-\,-\,-)$;  $\gamma^5 = i\gamma^0\gamma^1\gamma^2\gamma^3$, ${\sigma ^{\mu \nu }} = i[{\gamma ^\mu },{\gamma ^\nu }]/2$.}:
\begin{equation}
\label{l}
{\cal L} = {\cal L}_{\rm QED} + {\cal L}_{\rm T},
\end{equation} 
where
\begin{equation}
\label{qed}
{\cal L}_{\rm QED} = \bar\psi \left( \gamma ^{\mu }\left( i\partial _{\mu } + eA_{\mu }\right) - m\right)\psi - \frac{1}{4}F_{\mu \nu }F^{\mu \nu } - \frac{1}{2}{\left(\partial _{\mu}A^{\mu} \right)^2}
\end{equation}
is the Lagrangian of the standard QED in the Lorentz gauge, $\psi$ is the  electron-positron field, $m$ and $- e < 0$ are the electron mass and charge, $A^\mu$ and ${F^{\mu \nu }} = {\partial ^\mu }{A^\nu } - {\partial ^\nu }{A^\mu }$ are the 4-potential and the electromagnetic field strength tensor;
\begin{equation}
\label{t}
{\cal L}_{\rm T} = - \frac{1}{2}\bar \psi {\sigma ^{\mu \nu }}H_{\mu \nu}\psi
\end{equation}
is the Lagrangian of interaction with a constant tensor background field $H^{\mu\nu} = - H^{\nu\mu}$.

For the case of a quasielectric background field, we have  
 \[
H^{\mu \nu}H_{\mu \nu} < 0,\,H^{\mu \nu }{\widetilde H}_{\mu \nu} = 0, 
\]
where the dual tensor $\widetilde{H}_{\mu \nu } = \varepsilon _{\mu \nu \alpha \beta }H^{\alpha \beta }/2$
(note that for the quasimagnetic field $H^{\mu \nu}H_{\mu \nu} > 0$). 

Then there exists a special reference frame in which only the following components are different from zero (at the corresponding orientation of the coordinate axes): 
\begin{equation}
\label{qe}
H_{03} = - H_{30} = {\widetilde H}_{12} = - {\widetilde H}_{21} = b,
\end{equation}
so that the tensor background field in the considered case is equivalent to a 3-vector (we assume $b > 0$) 
\begin{equation}
\label{b}
{\bf b} = b{\bf e}_z, \, b = \left(- H^{\mu \nu}H_{\mu \nu}/2\right)^{1/2}.
\end{equation}

\section{Wave functions of an electron in the quasielectric background field}

The Dirac equation for the wave function of the electron in a tensor background field follows from (\ref{l})--(\ref{t}):
\begin{equation}
\label{dir}
\left(i\gamma ^\mu{\partial} _\mu - m- \frac{1}{2}{\sigma ^{\mu \nu }}H_{\mu \nu }\right)\psi  =0.
\end{equation}

Note that by substituting ${H_{\mu \nu }} \to {\mu _f}{F_{\mu \nu }}$ the equation (\ref{dir}) reduces to the Dirac--Pauli equation for a neutral fermion with magnetic moment $\mu _f$ in an external electromagnetic field $F_{\mu \nu }$. Therefore, we can use as solutions of the Dirac equation (\ref{dir}) in a background field the wave functions (with an appropriate change of notation) from a number of papers \cite{ter,kli,gav,ado} in which the effects associated with the motion of neutral fermions (in particular, neutrons) in electromagnetic fields were considered.

In the considered case of the quasielectric background field (\ref{qe}), let us represent Eq. (\ref{dir}) in the Hamiltonian form:
\begin{equation}
\label{hd}
\begin{split}
 i\frac{{\partial \psi }}{{\partial t}}  & = \widehat H\psi,\\
{\widehat H} & = {\boldsymbol{\alpha }} \cdot \widehat{\bf p}  + m\gamma^0  +ib\gamma^3,
\end{split}
\end{equation}
where the momentum operator $\widehat{\bf p} = - i\nabla$ and  $\boldsymbol{\alpha } = \gamma^0\boldsymbol{\gamma}$.

The solution of Eq. (\ref{hd}) has the form
\begin{equation}
\label{psi}
\begin{split}
\psi_{{\bf p}\zeta}(t,{\bf r})& = \frac{1}{\sqrt V}u\left( {{\bf{p}},\zeta } \right)\exp{\left(- iEt + i{\bf p} \cdot {\bf r} \right)},\\
u\left( {{\bf{p}},\zeta } \right)& = A
\begin{pmatrix}
Be^{-i\phi}\\
i\zeta B\\
i\zeta e^{-i\phi}\\
1
\end{pmatrix}.
\end{split}
\end{equation}
Here $V$ is the normalization volume;
\begin{equation}
\label{ab}
\begin{split}
A & = \frac{1}{2}\left(1 - \frac{m}{E}\right)^{1/2},\quad B = \frac{q}{E - m},\\
q & = p_{\bot} - \zeta(b - ip_z);
\end{split}
\end{equation}
 the spin quantum number $\zeta = \pm 1$, the electron energy
\begin{equation}
\label{en}
E = \sqrt {m^2 + |q|^2} = \sqrt{m^2 + p_z^2 + (p_{\bot} - \zeta b )^2}
\end{equation}
depends on $\zeta$, the longitudinal $p_z$ and transverse ${p_ \bot } = \sqrt {p_x^2 + p_y^2}$ components of the momentum ${\bf p}$; the angle $\phi$ in the expression for the wave function (\ref{psi}) defines the direction of the transverse momentum according to ${\bf p}_\bot = (p_x, p_y,0) = {p_ \bot }(\cos \phi,\sin \phi,0)$.

The wave function (\ref{psi}) describes the stationary state of the electron and is an eigenstate of the Hamiltonian $\widehat H$ (\ref{hd}), the momentum operator $\widehat{\bf p}$ and the spin operator \cite{stzhb}
\begin{equation}
\label{pi}
{\widehat \Pi } = {\gamma ^5}{\gamma ^\mu }{\widetilde H_{\mu \nu }}{p^\nu },
\end{equation}
where  $p^\nu = (E,{\bf p})$.
Taking into account Eqs. (\ref{qe}) and (\ref{b}), we represent the operator (\ref{pi}) as follows
\begin{equation}
\label{pi2}
{\widehat \Pi } = {\bf b}\cdot {\boldsymbol{\epsilon}} = b\epsilon_z = b\gamma^5 (\gamma^{1}p_y - \gamma^{2}p_x).
\end{equation}
Here the 3-vector of electric polarization  is introduced \cite{st}:
\begin{equation}
\label{eps}
{\boldsymbol{\epsilon}} = \gamma^{0}({\bf p}\times \boldsymbol{\Sigma}), \quad \boldsymbol{\Sigma} =\gamma^{5}\boldsymbol{\alpha}.
\end{equation}
The eigenvalues of the operator (\ref{pi2})  are related to the spin number $\zeta$ by the relation
\[
{\widehat \Pi }\psi_{{\bf p}\zeta} = \zeta b p_{\bot}\psi_{{\bf p}\zeta} . 
\]

\section{Radiative transition amplitude}

The amplitude of the transition of an electron $\left| i \right\rangle  = \left| {{\bf{p}},\zeta } \right\rangle  \to \left| f \right\rangle  = \left| {{\bf{p'}},\zeta '} \right\rangle$ with emission of a photon with 4-momentum $k^\mu = (\omega, \bf{k})$ and polarization vector $\bf{e}_{\lambda}$ is given, taking into account Eq. (\ref{psi}),  by the $S$-matrix element (see \cite{stzhb})
\begin{equation}
\label{sfi}
{S_{fi}} =  \frac{{ie}}{{\sqrt {2\omega V} }}\frac{{{{\left( {2\pi } \right)}^4}}}{V}\delta \left( {E' + \omega  - E} \right){\delta ^{(3)}}\left( {{\bf{p'}} + {\bf{k}} - {\bf{p}}} \right){\bf{e}}_{\lambda}^{*}\cdot \left\langle {\boldsymbol{\alpha }} \right\rangle,
\end{equation}
where
\begin{equation}
\label{alf}
\left\langle {\boldsymbol{\alpha }} \right\rangle  = {u^ + }\left( {{\bf{p'}},\zeta '} \right){\boldsymbol{\alpha }}u\left( {{\bf{p}},\zeta } \right).
\end{equation}

The frequency of the photon emitted in the direction ${\bf n} = {\bf k}/\omega$ ($\left| {\bf n} \right| = 1$) is determined by the conservation of energy and momentum (see delta functions in Eq. (\ref{sfi})):
\begin{equation}
\label{om}
\omega  = \frac{2bv_{\bot}}{1 - {\bf n}\cdot {\bf v}}\delta _{\zeta ',1}\delta _{\zeta , - 1}.
\end{equation}
Consequently, the radiation is due to the spin flip of the electron: $\zeta  =  - 1 \to \zeta ' = 1$.

Eq. (\ref{om}) is the leading term of the expansion in the parameter $b/m$, the smallness of which is provided by the present upper limit on the tensor background field strength \cite{sme3}:
\begin{equation}
\label{ub}
b \lesssim 10^{- 18}~{\rm eV}.
\end{equation}
In this approximation, which we will restrict below, ${\bf v} = {\bf p}/\varepsilon$ is the velocity of a free electron with energy $\varepsilon = E(b = 0) =\sqrt{m^2 + {\bf p}^2}$ (see Eq. (\ref{en})).

\section{Radiation power}

Using the general formulas of radiation theory \cite{stzhb} and the amplitude of the radiative transition (\ref{sfi}), we find the radiation power as
\begin{equation}
\label{wr}
W^{(\lambda)} = \frac{\alpha }{{2\pi }}\int {{d^3}k\delta \left( {E' + \omega  - E} \right)}\left|{\bf{e}}_{\lambda}^{*}\cdot \left\langle {\boldsymbol{\alpha }} \right\rangle\right|^{2},
\end{equation}
where the conservation of momentum is already taken into account: $\bf{p'} = \bf{p} - {\bf{k}}$.

Taking into account Eqs. (\ref{psi}), (\ref{ab}), (\ref{en}), and (\ref{om}), we obtain the matrix elements (\ref{alf}) and  $\left\langle{\alpha_0}\right\rangle$ ($\alpha_0 = I$ is the unit matrix) in the first order of expansion in  the background field $b$:
\begin{equation}
\label{1b}
\begin{split}
\begin{pmatrix}
\left\langle{\alpha_1}\right\rangle \\
\left\langle{\alpha_2}\right\rangle 
\end{pmatrix} & = \frac{1}{2\varepsilon p_{\bot}^2}\left[-i\omega\varepsilon + (p_{\bot} - i p_z)\left(\frac{i\omega p_{\bot}}{\varepsilon - m} - k_z\right)\right]
\begin{pmatrix}p_y\\- p_x\end{pmatrix},\\
\begin{pmatrix}
\left\langle{\alpha_3}\right\rangle \\
\left\langle{\alpha_0}\right\rangle 
\end{pmatrix} & = \frac{k_xp_y - k_yp_x}{2\varepsilon p_{\bot}^2}\begin{pmatrix} p_{\bot} - ip_z\\
\displaystyle im + \frac{p_z(p_{\bot} - ip_z)}{\varepsilon - m}\end{pmatrix}.
\end{split}
\end{equation}
Note that the matrix elements (\ref{1b}) satisfy the relation
\begin{equation}
\label{emc}
\omega\left\langle {{\alpha_0}} \right\rangle - {\bf{k}} \cdot \left\langle {\boldsymbol{\alpha }} \right\rangle  = 0,
\end{equation}
which follows from the conservation of the electromagnetic current \cite{ll}.

Putting ${d^3}k = {\omega ^2}d\omega d\Omega$ in Eq. (\ref{wr}), where $d\Omega$ is the solid angle element in the direction $\bf n$, we integrate over $\omega$ using the delta function. As a result, we find the angular distribution of the radiation power: 
\begin{equation}
\label{wl}
\frac{{dW^{(\lambda)}}}{{d\Omega }} = \frac{\alpha }{{2\pi }}\frac{{{\omega ^2}}}{{1 - {\bf{n}} \cdot {\bf{v}}}}\left|{\bf{e}}_{\lambda}^{*}\cdot \left\langle {\boldsymbol{\alpha }} \right\rangle\right|^{2},
\end{equation}
where the frequency $\omega$ is defined in Eq. (\ref{om}).

Summation in Eq. (\ref{wr}) by polarizations according to the known formula \cite{stzhb}
\begin{equation}
\label{ps}
\sum\limits_\lambda {e_\lambda ^{ * i}e_\lambda ^k} = {\delta ^{ik}} - {n^i}{n^k}
\end{equation}
yields in view of (\ref{emc})
\begin{equation}
\label{wrs}
\frac{dW}{d\Omega} = \frac{\alpha }{2\pi }\frac{\omega ^2}{1 - {\bf n} \cdot {\bf v}}\left(\left|\left\langle\boldsymbol{\alpha}\right\rangle\right|^2 - \left|\langle\alpha _0\rangle\right|^2\right).
\end{equation}

Taking into account Eqs. (\ref{om}) and (\ref{1b}), as well as the axial symmetry of the background field (\ref{b}), which allows us to set ${\bf p} = (p_x,0,p_z)$, we represent Eq. (\ref{wrs}) in the form
\begin{equation}
\label{wn}
\frac{dW}{d\Omega} =W_0\frac{(1 - v^2)v_x^2}{(1 - n_xv_x - n_zv_z)^5}\left[(1 - n_zv_z)^2 - n_x^2v_x^2 - (1 - v^2)n_y^2\right],
\end{equation}
where
\begin{equation}
\label{w0}
W_0 = \frac{2\alpha}{\pi}\frac{b^4}{m^2}.
\end{equation}
In the spherical coordinate system with the polar axis $Oz$, we have in Eq. (\ref{wn})
\begin{equation}
\label{n}
\begin{split}
d\Omega & = \sin\theta d\theta d\varphi,\\
{n_x} & = \sin \theta \cos \varphi ,\quad {n_y} = \sin \theta \sin \varphi ,\quad {n_z} = \cos \theta.
\end{split}
\end{equation}

To calculate the total radiation power, it is convenient to express the angles in Eq. (\ref{n}) through the angles (marked by the index 0) in a reference frame moving with the velocity $v_z$ along the $Oz$ axis (as in synchrotron radiation theory \cite{st}), using an appropriate boost that does not change the original configuration of the background field (\ref{b}):
\begin{equation}
\label{LT}
\begin{split}
\begin{pmatrix}n_{x}\\n_{y}\end{pmatrix} &=
\frac{\sqrt {1 - v_z^2}}{1 + v_{z}n_{0z}}
\begin{pmatrix}n_{0x}\\n_{0y}\end{pmatrix},\quad
n_z = \frac{n_{0z}+ v_z}{1 + v_{z}n_{0z}},\\
v_x & = v_{0x}\sqrt {1 - v_z^2}, \quad d\Omega  = \frac{1 - v_z^2}{(1 + v_{z}n_{0z})^2}d\Omega _0.
\end{split}
\end{equation}
Using Eq. (\ref{LT}), we represent Eq. (\ref{wn} as
\begin{equation}
\label{dw0}
\frac{dW}{d\Omega_0} =W_0(1 + v_zn_{0z})\frac{v_0^2(1 - v_0^2)}{(1 - v_0n_{0x})^5}\left[1 - v_0^2n_{0x}^2 - (1- v_0^2)n_{0y}^2\right].
\end{equation}
Here $v_0 \equiv v_{0x}$ is invariant with respect to boosts along the $Oz$ axis:
\begin{equation}
\label{v0}
v_0 = \frac{v_{\bot}}{\sqrt{1 - v_z^2}}, \quad v_{\bot} = \sqrt{v^2 - v_z^2}.
\end{equation}

The integration of the angular distribution (\ref{dw0}) is simplified if we choose $Ox$ as the polar axis. Then $n_{0x} = \cos \alpha ,n_{0y} = \sin \alpha \cos \beta ,n_{0z} = \sin \alpha \sin \beta$, which allows us to perform independent integration over the angles $\alpha$ and $\beta$, and as a result we obtain the total radiation power
\begin{equation}
\label{wtot}
W = \frac{8\pi}{3}W_0\frac{v_0^2(1 + 3v_0^2)}{(1 - v_0^2)^2} = \frac{16}{3}\frac{\alpha b^4}{m^2}t^2(1 + 4t^2),
\end{equation}
where 
\begin{equation}
\label{t}
t = \gamma v_{\bot} = \frac{p_{\bot}}{m}
\end{equation}
with the Lorentz factor $\gamma = {\varepsilon}/m = 1/\sqrt{1 - v^2}$.

For the unpolarized electron it is necessary to introduce an additional factor $1/2$  in the right side of Eq. (\ref{wtot}).
Note that the power (\ref{wtot})  is a Lorentz invariant (see, e. g., \cite{stzhb}), which explains its dependence only on $v_0$. 

\section{Polarization of the radiation}

To describe the polarization of radiation, we introduce, as in the theory of synchrotron radiation, the $\sigma$- and $\pi$-components of the linear polarization (see \cite{stzhb}):
\begin{equation}
\label{pol}
{\bf e}_\sigma = \frac{{\bf e}_z \times{\bf n}}{\left|{\bf e}_z \times{\bf n}\right|} = \frac{- n_y{\bf e}_x + n_x{\bf e}_y}{\sqrt{1 - n_z^2}},\quad
{\bf e}_\pi = {\bf n} \times {\bf e}_\sigma = \frac{ {\bf e}_z - n_z{\bf n}}{\sqrt{1 - n_z^2}}.
\end{equation}

From (\ref{wr}), (\ref{1b}), (\ref{LT}) and  (\ref{pol}), we find angular distributions of radiation powers of linear polarization components:
\begin{equation}
\label{dwp}
\begin{pmatrix}dW^{\sigma}/d\Omega_0 \\dW^{\pi}/d\Omega_0\end{pmatrix} = W_0(1 + v_zn_{0z})\frac{v_0^2(1 - v_0^2)}{(1 - v_0n_{0x})^5}\frac{1}{1 - n_{0z}^2}
\begin{pmatrix}n_{0x}^2\left[1 - v_0^2(1 - n_{0z}^2)\right]\\n_{0y}^2\left[v_0^2 + (1- v_0^2) n_{0z}^2)\right]\end{pmatrix}.
\end{equation}
Their sum gives (\ref{dw0}), as it should be. 

By integrating the distributions (\ref{dwp}) over the angles, we obtain the total radiation powers of the polarization components
\begin{equation}
\label{wp}
\begin{pmatrix}W^{(\sigma)}\\W^{(\pi)}\end{pmatrix} = \frac{\pi}{3}W_0\frac{v_0^2}{(1 - v_0^2)^2}\begin{pmatrix}(6 + 23v_0^2 - v_0^4)\\
(2 + v_0^2 + v_0^4)\end{pmatrix},
\end{equation}
the sum of which coincides with Eq. (\ref{wtot}).

It follows from (\ref{wp}) that the radiation is linearly polarized ($\sigma$ component predominates), and the degree of polarization is 
\begin{equation}
\label{dp}
P = \frac{W^{(\sigma)} - W^{(\pi)}}{W} = \frac{2 + 11 v_0^2 - v_0^4}{4(1 + 3 v_0^2)} = \frac{2 + 15t^2 + 12t^4}{4(1 + t^2)(1 + 4t^2)}.
\end{equation}
It increases monotonically with velocity, from $1/2$ at $v_0 \to 0 (t \to 0)$ to $3/4$ at $v_0\to 1(t\to\infty)$.

\section{Probability of radiative transition}
The angular probability distribution of radiation is obtained by multiplying the right-hand side of Eq. (\ref{dw0}) by $1/\omega$ and taking into account Eqs. (\ref{om}), (\ref{w0}), (\ref{LT}), and (\ref{v0}):
\begin{equation}
\label{dpr}
\frac{dw}{d\Omega_0} = \frac{\alpha b^3}{\pi m^2}\sqrt{1 - v_z^2}\frac{v_0(1 - v_0^2)}{(1 - v_0n_{0x})^4}\left[1 -v_0^2n_{0x}^2 - (1 - v_0^2)n_{0y}^2\right].
\end{equation}
Integration of the distribution (\ref{dpr}) over the angles gives the total probability of radiation
\begin{equation}
\label{pr}
\begin{split}
w &= \sqrt{1 - v_z^2}w^{(0)} = w_{0}v_{\bot}(1 + 3t^2),\\
w^{(0)} &= w_0\frac{v_0(1 + 2v_0^2)}{1 - v_0^2}, \quad w_0 = \frac{8}{3}\frac{\alpha b^3}{m^2}.
\end{split}
\end{equation}
Here, the characteristic dependence on the longitudinal velocity $v_z$ expresses the law of probability transformation at boost according to the special relativity. 

Consider the average radiative energy loss of an electron, i. e. the average photon energy:
\begin{equation}
\label{av}
\langle\omega\rangle = \frac{\int{\!\omega}dw}{\int{\!dw}} = \frac{W}{w}.
\end{equation}
From Eq. (\ref{av}), taking into account Eqs. (\ref{wtot}) and (\ref{pr}), we find
\begin{equation}
\label{av2}
\langle\omega\rangle = 2b{\gamma}t\frac{1 + 4t^2}{1 + 3t^2},
\end{equation}
and the maximum photon energy (see Eq. (\ref{om}))
\begin{equation}
\omega_{\rm max} = 2b{\gamma}t(1 + v).
\end{equation}

According to Eq. (\ref{av}), on average after a time interval
\begin{equation}
\label{tr}
\tau_R = 1/w
\end{equation}
the electron emits a photon, having made a spin-flip transition to a radiatively stable state (see Eqs. (\ref{en}) and (\ref{om})). 
Therefore, the initially unpolarized electron beam becomes fully polarized, with $\tau_R$ being the characteristic polarization time.
Using Eqs. (\ref{pr}) and (\ref{tr}), we find the radiative polarization length
\begin{equation}
\label{lr}
L_R = v\tau_R = \frac{v}{w_0 v_{\bot}(1 + 3t^2)}.
\end{equation}

\section{Discussion}
Let us compare the results obtained for the background quasielectric field (type E) with the corresponding results for the quasimagnetic field (type M) \cite{bor2,bor3,bor4}.

In the region of small velocities they are very different: for type E the power and probability of radiation go to zero at $v \to 0$ (see Eqs. (\ref{wtot}) and (\ref{pr})), while for type M even a resting electron emits a photon. This difference is explained by the fact that in the case of type M the contribution of the magnetic moment of the electron dominates \cite{bor1}, while for type E the radiation is due to the interaction of the induced electric moment (at $v > 0$) with the quasielectric field (see Eqs. (\ref{pi2}) and (\ref{eps})).

At $v = v_{\bot} \to 1$ (the high energy region, $\gamma \gg 1$), the results for types E and M practically coincide. This is explained by the fact that at the Lorentz transformation to the rest frame of the electron both configurations of the background field become practically indistinguishable from the configuration of the so-called crossed field  (cf. \cite{ll,rit}). In this frame, the quasielectric ${\bf b}_0$ and quasimagnetic ${\bf h}_0$ fields are orthogonal to each other and are equal in magnitude, as it follows from the explicit form of the corresponding transformations (see, e.g., \cite{jj}):
\[
{\bf b}_0 = \gamma {\bf b}, \, {\bf h}_0 = - \gamma {\bf v}\times {\bf b}\quad  {\rm for} \,\, {\bf v}\cdot {\bf b} = 0,
\]
and
\[
{\bf h}_0 = \gamma {\bf h}, \, {\bf b}_0 =  \gamma {\bf v}\times {\bf h} \quad {\rm for} \,\, {\bf v}\cdot {\bf h} = 0,
\]
respectively.

It follows from Eqs. (\ref{wtot}), (\ref{pr}), (\ref{av2}), and (\ref{lr}) that the effects of Lorentz violation increase with increasing electron energy, being more noticeable for the transverse motion of the electron ($\bf{v} \bot \bf{b}$). 

However, our results are valid under the condition that the recoil in the emission of a photon by an electron is small, i.e. 
\[
\chi = \frac{\langle\omega\rangle}{\varepsilon} \ll 1.
\]
Hence, taking into account Eqs. (\ref{av2}) and {\ref{t}), in case $v_{\bot} \to 1$ we obtain the constraint
\begin{equation}
\label{rec}
\chi \simeq \frac{\gamma b}{m} \ll 1.
\end{equation}
In the same case for the radiative polarization length (\ref{lr}) we obtain (in usual units)
\begin{equation}
\label{lr2}
L_R \simeq \frac{\lambdabar_e}{8\alpha}\left(\frac{m}{b}\right)^{3}\gamma^{- 2},
\end{equation}
where $\lambdabar_e$ is the Compton wavelength of the electron.

For numerical estimations, we set $b = 5\times 10^{- 18}~{\rm eV}$ (see Eq. (\ref{ub})) and 
$\varepsilon = 10^{16}~{\rm GeV}$ (the energy scale of the Grand Unification \cite{lan,nag}).
Then from Eqs. (\ref{rec}) and (\ref{lr2}) we obtain $\chi \simeq 2\times 10^{- 4}$, $\langle\omega\rangle = 2\times 10^{12}~{\rm GeV}$ (cf. with the maximum registered energy of particles in cosmic rays $\simeq 10^{11}~{\rm GeV}$ (see the review \cite{anc})), and $L_R \simeq 2\times 10^{21}~{\rm cm}$
 (for comparison, the distance from the Sun to the nearest star $\simeq 4\times 10^{18}~{\rm cm}$ \cite{ns}, 
and the distance from the Sun to the center of the Galaxy $\simeq 2.5\times 10^{22}~{\rm cm}$ \cite{pdg}).

\section{Conclusion}

In the framework of the Standard Model Extension, we have calculated the power and probability of electromagnetic radiation from an electron 
in a constant quasielectric background field, simulating a Lorentz-violating vacuum. We found that the radiation has a significant linear polarization,
which reaches 75 \% for high energy electrons. It is shown that the radiative transition due to spin flip leads to a complete polarization of the initially 
unpolarized electron beam along the direction of the background field. The obtained results are compared with the results of our previous works 
for the background field of the quasimagnetic type. We have shown that the considered radiative effect can be observed under astrophysical conditions 
for ultrahigh energy electrons.

\section*{Acknowledgment}

The author thanks K. V. Stepanyantz for useful discussions.

\let\doi\relax


\end{document}